\documentclass[12pt]{iopart}
\usepackage{graphicx}
\usepackage{refstyle}
\usepackage{multirow}
\usepackage[english]{babel}
\usepackage{xcolor}

\usepackage{iopams}
\expandafter\let\csname equation*\endcsname\relax
\expandafter\let\csname endequation*\endcsname\relax
\usepackage[utf8]{inputenc}

\begin{document}

\title{Turbulence-induced transport dynamo mechanism }

\author{ Chang-Mo Ryu $^{1,2}$}

\address{$^1$ Department of Physics, POSTECH, Pohang, South Korea}
\address{$^2$ Center for Relativistic Laser Science, IBS, Gwangju, Korea  }
\ead{ryu201@postech.ac.kr}
\vspace{10pt}
\begin{abstract}
The transport dynamo mechanism, which describes the magnetic field generation by diffusion flow is reviewed. In this mechanism, the cross-field transport
caused by the random motion of fluid breaks the frozen-flux approximation, and the resulting cross-field diffusion that can generate the magnetic field. Turbulence can play an
important role in inducing such random motion. Compared to the conventional dynamo mechanism, this transport mechanism has several special features that the field generation can occur on a very slow time scale because the mechanism is mediated by
diffusion and that this mechanism is practically meaningful only when there is density inhomogeneity. Turbulence can significantly enhance cross-field diffusion far beyond collisional transport. The physical meanings of the diffusion-generated magnetic fields are discussed in detail.
 \\
\end{abstract}

\section{Introduction}

The dynamo mechanism deals with the generation of magnetic fields  by fluid motion. 
Identifying the large-scale magnetic field generation mechanism  is crucial  in  laboratory and astrophysical plasmas \cite{Kulsrud}.  
The current theoretical structure of the dynamo mechanism  started from Cowling's anti-dynamo theorem.   Cowling has demonstrated that the generation of the  magnetic field by  simple two-dimensional flows is not possible \cite{Cowling}. Since then, the  dynamo theory has been  advanced to find a fluid motion to generate the magnetic field in  complicated three-dimensional geometry, which has a long history of development \cite{Brandenburg18}.
In his monograph, Cowling  noted  \cite{Cowling}  that
  the real difficulty of the dynamo problem is 'finding a flow velocity relative to the field lines, not the total velocity of the material,  where the field lines are carried about with the velocity of the material'.   The flow velocities required to sustain the  magnetic field in the Sun or the Earth are  tiny compared to other velocities in the same material, but they have to be the  relative ones to the field lines.  There can exist many flow velocities much greater than the required velocities in a conducting medium, but the flow in such a medium tends to move together with the field line. Consequently, overcoming the  effect of a nearly frozen-in field  has become extremely difficult in a  highly conducting medium. Thus, according to Cowling's viewpoint, the essence of any dynamo theory may lie in finding  a relative flow velocity  to the magnetic field lines under the  condition that fields are nearly frozen into the fluid.

From this perspective,  cross-field diffusion in plasmas  is of  interest.   By definition, cross-field diffusion   describes the plasma transport across the magnetic field,  where fluid elements   can cross the magnetic field lines via random motion.  Collisions, turbulence, and instabilities can contribute to such random motion. The cross-field plasma transport is important  in  thermonuclear fusion plasmas so that it has been intensively investigated.  And yet, its potential ability to generate the magnetic field has been mostly overlooked.


 

The possibility of a diffusion flow to generate the magnetic field  was  proposed by the author along with his co-authors  \cite{Ryu_LA, Ryu_Scripta, Ryu_PoP, Ryu_JPD, Ryu_EPS2019}. In this paper,   the  essential physics mechanism is elucidated based on previous work,  to help the reader who might be interested in this new mechanism of magnetic field generation.

\section{Alfvén's  frozen-in magnetic field theorem}

Hannes Alfvén put forward the idea for the first time in 1942 that "the matter of liquid" is 'fastened' to the lines of force" in a conducting medium \cite{Alfven1}. 
This idea  can be understood as follows

The basic set of  equations describing magnetohydrodynamics (MHD) in a conducting medium can be written in cgs units as,

the continuity equation
\begin{equation}
\frac{\partial \rho}{\partial t}+\nabla \cdot(\rho \mathbf{v})=0, \label{1}
\end{equation}
the momentum equation
\begin{equation}
\rho\left(\frac{\partial \mathbf{v}}{\partial t}+(\mathbf{v} \cdot \nabla) \mathbf{v}\right)=-\nabla p+\frac{1}{c} (\mathbf{J} \times \mathbf{B}), \label{2}
\end{equation}
 Ohm's law  for  MHD
{\begin{equation}
\mathbf{E}+ \frac {1}{c} \mathbf{v} \times \mathbf{B}  =   {\eta} \mathbf{J}, \label{3}
\end{equation}
where $c$ is the speed of  light and  $\eta$ is the resistivity.

These equations need to be supplemented by
Faraday's law 
{\begin{equation}
\frac{1}{c} \frac{\partial \mathbf{B}}{  \partial t}=-\nabla \times \mathbf{E} , \label{4}
\end{equation}
and Ampere's law
\begin{equation}
 \mathbf{J}=\frac{c}{4 \pi} (\nabla \times \mathbf{B}), \label{5}
\end{equation}
 and then  Ohm's law can give rise to  the magnetic induction equation as
{\begin{equation}
\frac{\partial \mathbf{B}}{\partial t}-\nabla \times(\mathbf{v} \times \mathbf{B}) =   \frac{\eta c^{2}}{4 \pi} \nabla^{2} \mathbf{B}. \label{6}
\end{equation}

For a fluid with infinite electric conductivity, ${\displaystyle \eta \rightarrow 0}$, which is called the ideal MHD fluid, the final term in the induction equation vanishes, and
the magnetic induction equation simplifies to  

\begin{equation}
{\partial {\mathbf {B}} \over \partial t}={\mathbf {\nabla }}\times ({\mathbf {v}}\times {\mathbf {B}}). \label{7}
\end{equation}

The ideal MHD  approximation is very useful describing  many plasmas such as tokamaks, stars, galaxies, and accretion disks.  In many astrophysical plasmas, ideal MHD corresponds to a substantial magnetic Reynolds number.

When a  closed curve ${C}$  is drawn in plasma   
 the magnetic flux on the surface ${{\mathbf {S}}}$  enclosed in this curve  is defined by
\begin{equation}
\Phi_B=\oint_{S} \mathbf{B} \cdot d \mathbf{S}, \label{8}
\end{equation}
yields the time derivative of the magnetic flux as
\begin{equation}
\frac{d \Phi_{B}}{d t}=\int_{S} \frac{\partial \mathbf{B}}{\partial t} \cdot d \mathbf{S}+\oint_{C} \mathbf{B} \cdot \mathbf{v} \times d \mathbf{l}. \label{9}
\end{equation}
Here  ${ d{\mathbf {l}}}$  is the differential line element of curve $C$. 
Curve $C$  moves with the fluid at velocity $\mathbf{v}$.   
Then, by using the magnetic induction equation (\ref{7}), it is possible to get the time derivative of the magnetic flux as

\begin{equation}
\frac{d \Phi_{B}}{d t}=\int_{S} \mathbf{\nabla} \times(\mathbf{v} \times \mathbf{B}) \cdot d \mathbf{S}+\oint_{C} \mathbf{B} \cdot \mathbf{v} \times d \mathbf{l}. \label{10}
\end{equation}

 The first term on the right-hand side (RHS) can be written as, by  Stoke's theorem, 
\begin{equation}
\int_{S} \mathbf{\nabla} \times(\mathbf{v} \times \mathbf{B}) \cdot d \mathbf{S}=\oint_{C} \mathbf{v} \times \mathbf{B} \cdot d \mathbf{l},  \label{11}
\end{equation}

and as a result, the first term cancels the second term on the RHS,  making the total  time derivative of the  magnetic flux vanish as,

\begin{equation}
\frac{d \Phi_{B}}{d t}= \oint_{C} \mathbf{v} \times \mathbf{B} \cdot d \mathbf{l}+\oint_{C} \mathbf{B} \cdot \mathbf{v} \times d \mathbf{l} =0. \label{12}
\end{equation}

Therefore,  in a conducting medium with infinite conductivity  the magnetic flux is conserved. 
The conservation of magnetic flux allows flux tube be defined  as a  cylindrical volume of magnetic field lines embedded in fluid, where the fluid moves along with the field lines.

 The frozen-in-field is one of the  most important concepts in MHD.    In a fully ionized plasma, electrons and ions revolve around the magnetic field line  with the center of  rotation in the magnetic field line. The average mass of a rotating electron or an ion is located at its center of rotation. When the magnetic field lines move, fluid elements consisting of electrons and ions move along with the magnetic field line, as if the mass were frozen in the magnetic field line.  Many important macroscopic behaviors of a plasma can be easily understood by using this simple concept.
The frozen-in-field is a very useful concept in MHD, and its validity is rarely challenged despite the fact that various approximations are adopted in MHD. 

However, in a dynamic many-body system such as plasma, underlying microscopic motions can alter the principles of MHD. That is, random motion can break the frozen-in-field condition.  More specifically, random motion caused by collisions, electromagnetic fluctuations, instabilities, and turbulence can distort the motions of electrons and ions, and  break the plasma frozen in  magnetic fields.  Diffusion across a magnetic field can occur by random motion. In fact, cross-field diffusion is a well observed phenomenon in laboratory plasmas.   Then, the  question  arises whether the  velocity associated with diffusion, intrinsically relative to the magnetic field line,  can generate the magnetic field.  This question is the main subject addressed  in this paper.


\section{Can a magnetic field be generated by diffusion?}

Diffusion can be  described by Fick's law,

\begin{equation}
{\mathbf {\Gamma}} =-D \nabla n = n {\mathbf {v_D}}, \label{t-1}
\end{equation}

where $n$ is the density, $D$  the diffusion coefficient, and ${\mathbf {v_D}}$ is  the diffusive flow velocity.

The diffusion flow $\mathbf{\Gamma}$ across the magnetic field, denoted by $n\mathbf{v_D}$,  can induce the electric field through the relationship 
{\begin{equation}
\Delta \mathbf{E}=  -\frac{1}{c} ( \mathbf{v_D} \times \mathbf{B})= -\frac{1}{nc} \mathbf{\Gamma} \times \mathbf{B} . \label{t-2}
\end{equation}

If this induced electric induced  is added to  equation (\ref{3}), Ohm's law becomes, 
{\begin{equation}
\mathbf{E}+\frac{1}{c} \mathbf{v} \times \mathbf{B}  =  - \frac{1}{nc} \mathbf{\Gamma} \times \mathbf{B}+ \eta \mathbf{J}. \label{t-3}
\end{equation}

Taking the curl  and multiplying $c$ of this equation yields

{\begin{equation}
\frac{\partial \mathbf{B}}{\partial t}-\nabla \times(\mathbf{v} \times \mathbf{B}) =
 \nabla \times ( {\frac{1}{n} \mathbf{\Gamma} \times \mathbf{B}}) +
  \frac{\eta c^{2}}{4 \pi} \nabla^{2} \mathbf{B}, \label{t-4}
\end{equation}
which is the magnetic induction equation that includes the diffusion effect.
It should be noted  here that the transport effect is different from the thermoelectric current dynamo proposed by Haines \cite{Haines}, in which thermal electron transport occurs due to the temperature difference.

  Diffusion flow can carry charge, mass, momentum, and heat, but the diffusion flow cannot simply  be replaced with an advection flow.   Diffusion is an irreversible process, and  thus, the advection flow cannot fully represent the nature of diffusion. For instance,  diffusion flow may not induce oscillatory behavior, and  the  magnetic field may not pull back a diffusing fluid  that passes through the field line because the diffusion process is not reversible.    Therefore, great care must be taken when expressing the transport of diffusion as MHD flow. However,  transport activity can induce the electric field that satisfies Ohm's law  given in (\ref{t-3}).}



The magnetic field induced by the relative velocity of diffusion can be  more clearly understood by considering  it in a flux tube. At the boundaries of a flux tube,  curve $C$ can be drawn fixed to the magnetic field  lines. Tube boundaries can move with advection, but  diffusion can permeate through the magnetic boundaries.  This fixed curve selection to the magnetic field lines is different from the one moving with fluid typically adopted.  When flux conservation is obtained in equation (\ref{12}), the curve was  chosen  to move with  fluid, i.e., in  Lagrangian coordinates. However, in practice,  magnetic fields are measured at  fixed positions in an Eulerian manner rather than following fluid.  It will be very difficult or even impossible to measure the magnetic field in the frame of diffusing fluid elements.  In other words,  measuring the  magnetic field in Eulerian coordinates is just a practical method.
  Then, on the surface enclosed  in a curve, $C$ fastened to the magnetic field lines, the time derivative of the flux $\Phi  =  \int  \boldsymbol{B}  \cdot \mathrm{d} \boldsymbol{S}$  becomes in the ideal MHD limit as,  from equation  (\ref{t-4}), 
\begin{eqnarray}
	{\frac {d \Phi} {d t}}  
	&=&  \oint_{C} \mathbf{(v}+ {\frac{1}{n} \mathbf{\Gamma}      )} \times \mathbf{B}  \cdot d \mathbf{l} +\oint_{C} \mathbf{B} \cdot \mathbf{v} \times d \mathbf{l} \nonumber,  \label{t-7-1} \\
	&=&  \oint_{C} {\frac{1}{n} \mathbf{\Gamma}  } \times \mathbf{B}  \cdot d \mathbf{l}  , \nonumber  \label{t-7-2-0} \\
	& =& - 2 \pi r v_{D} B_{z} ,  \label{t-7-2}
\end{eqnarray}
with $v_{D}$ defined by
\begin{equation}
 v_{D}=-D \frac{1}{n}  \frac { \partial n}{ \partial { r}}.  \nonumber  
\end{equation}
 
This result indicates that  the magnetic flux contained in the flux tube can increase or decrease depending on the  direction of diffusion flow.  In a uniform plasma, there is no such directional transport and thus, there is no flux change.
For a centrally high-density profile, such as a Gaussian profile, the magnetic flux decreases, but  for a centrally low-density profile such as a ring-shaped plasma, the magnetic flux can increase.

When another curve $C'$ is drawn to follow  diffusion, the time derivative of the flux vanishes,
\begin{equation}
	{\frac {d \Phi} {d t}}  
	=  \oint_{C'} \mathbf{(v}+ {\frac{1}{n} \mathbf{\Gamma}      )} \times \mathbf{B}  \cdot d \mathbf{l} +\oint_{C'} \mathbf{B} \cdot (\mathbf{v}+\mathbf{v_D}) \times d \mathbf{l} = 0, \\ \label{t-9-1}
\end{equation}
The magnetic flux on the surface enclosed in $C'$  is conserved. 
According to Alfvén's theorem, if the flux is conserved, the magnetic field lines  move with the fluid. 
 
\begin{equation}
\mathbf{B} \times [ \nabla \times ( \mathbf{E} \times  \frac{1}{c} ( \mathbf{v} \times \mathbf{B}  ))   ] = 0 . \label{9-1-8}
\end{equation}

The diffusive flow velocity  $\mathbf{v_D}$ can satisfy the condition of  Newcomb \cite{Falthammer, Newcomb}.  This result allows  fluid to move covariantly while maintaining a common magnetic field line.
 
 Equation (\ref{9}) for flux conservation can be rewritten using Ohm's law as
\begin{eqnarray}
\frac{d \Phi_{B}}{d t} &= &\int_{S} \frac{\partial \mathbf{B}}{\partial t} \cdot d \mathbf{S}+\oint_{C} \mathbf{B} \cdot \mathbf{v} \times d \mathbf{l} \label{9-1} \nonumber \\
   &=& -c \int_{S}  \nabla \times \big[ \mathbf{E}   + \frac{1}{c}  (\mathbf{v} \times \mathbf{B}) \big] \cdot d \mathbf{S}    = 0.     \label{9-1-7}
 \end{eqnarray}
  This result  shows that if Ohm's law is true, flux conservation  is valid.   For the diffusion flow, the relationship between Ohm's law and the flux conservation still holds.  Many approximations are used in deriving MHD equations. Alfvén's theorem relies on these approximations and has intrinsic ambiguity at the microscopic level \cite{Bellan}.    For diffusion flow, Ohm's law can ensure flux conservation. Therefore working directly with Ohm's law (or equations of magnetic induction) can be one way to avoid the conceptual ambiguity associated with frozen-in magnetic fields or the motion of magnetic field lines.   Ohm's law can gives a more accurate and clearer picture on field generation than flux conservation.

\section{Transport dynamo by Bohm diffusion}

The diffusion coefficient  $D$ is determined by  collisions, instability,   and  turbulence.  Plasma diffusion is caused by underlying electric and  magnetic fluctuations. In a fusion plasma device, anomalous diffusion  can result in  significant  plasma transport towards the wall across the confining magnetic field. 

In a fully ionized plasma,
the plasma diffusion coefficient $D$ can be taken in the  range from  $D_{\mathrm{class }}$ to $D_{\mathrm{Bohm}}$,  where  $D_{\mathrm{class }}$ is the classical diffusion coefficient due to binary collisions and $D_{\mathrm{Bohm}}$ is the anomalous diffusion first discovered  by Bohm \cite{Bohm, Spitzer}.  
The classical and Bohm diffusion coefficients can be written as \cite{Chen, Dawson},

\begin{equation}
D_\mathrm{class} = \eta \frac{nk(T_i+T_e)c^2}{B^2} = \eta \frac{2nkT c^2}{B^2}, \label{t-5}
\end{equation}

and
\begin{equation}
 D_{\mathrm{Bohm}} = \frac{kT_i c}{16eB}   = \frac{kTc}{16eB}. \label{t-6}
\end{equation}

 For  electron, ion, and total temperatures, $T_e=T_i=T$  is taken. The Bohm diffusion coefficient has a numerical factor uncertainty of 3 to 4, which is related to the underlying instability. 

  It was evidence in many experiments  that  the diffusion across the magnetic field  is of  Bohm-type,  which implies that the underlying  electromagnetic fluctuations are important.  
Therefore,  the Bohm diffusion can be  taken as the maximum diffusion coefficient.    
For a highly turbulent plasma developed by  fluid instabilities, such as a Kelvin-Helmholtz instability,  $D$ can be estimated from the statistical mean of the square of velocity fluctuations divided by the correlation time,  as  $D= <v^2> \tau$, where $v$ is the fluctuation velocity and  $\tau$ is the correlation time \cite{Dawson}.

The growth rate of the magnetic field  can be  estimated in the order of magnitude  using  equation (\ref{t-4}).  Assuming that the advection flow velocity is zero,  $\mathbf{v}=0$,  the magnetic induction equation becomes

\begin{equation}
{\gamma{\mathbf {B}} } \sim - \frac{1}{{L }} ({ {v_D}} {\mathbf {B}})-    \frac{\eta c^{2}}{4 \pi} \frac{1}{L^2} \mathbf{B}.\label{t-10}
\end{equation}

The rate of change of the magnetic field can be, then, found as

\begin{equation}
{\gamma } \sim  - \frac {v_D}{L} -    \frac{\eta c^{2}}{4 \pi} \frac{1}{L^2} . \label{t-11}
\end{equation}


 According to equation (\ref{t-1}),  the magnitude of the diffusive flow velocity  is in the order of 
\begin{equation}
 v_D  \sim  \frac{D}{L}. \label{t-12}
\end{equation}

Thus, for classical diffusion, 


\begin{eqnarray}
{\gamma } &\sim& - \frac{D_{class}}{L^2}  -    \frac{\eta c^{2}}{4 \pi} \frac{1}{L^2} , \label{t-13} \nonumber \\
          & \sim &  - \beta \frac{ \eta c^2}{4\pi L^2}  -    \frac{\eta c^{2}}{4 \pi} \frac{1}{L^2} ,  \label{t-14}
\end{eqnarray}
using the plasma beta  $\beta = 8 \pi \frac{nKT}{B^2}$.

 The first term in equation (\ref{t-14}) is comparable to the second term when the plasma beta is  in the order of $\mathcal{O}(1)$. When
the plasma pressure is about the magnetic energy $P(=2nkT) \sim B^2/{8\pi}$,   $D_\mathrm{class}  \sim \eta c^2 / 4\pi$, and the effects of classical diffusion becomes comparable to those by resistive dissipation.    Consequently, in the case of an outward flow, the effective resistivity is doubled by classical diffusion, but in the case of an inward flow, the magnetic field can be maintained  against resistive dissipation.

For the Bohm diffusion normally caused by  turbulence or instability, the magnetic induction term becomes much greater than the resistive dissipation term. The magnetic induction equation becomes roughly

\begin{equation}
\gamma {\mathbf{B}}  \sim \frac{1}{L} {{v_D}} \mathbf{B}
\sim \frac{1}{16L^2} \frac{kTc}{e B} {\mathbf{B}}, \label{t-15}
\end{equation}
 which yields  the magnetic amplification rate for an inward diffusion as 
\begin{equation}
\gamma \sim \frac{1}{16L^{2}} \frac{kTc}{ e B}. \label{t-16}
\end{equation}

The major parameters in this growth rate estimation are the length scale $L$, magnetic field $B$, and temperature $T$. The growth rate decreases with the magnetic field and length scale, but increases with temperature.  The growth rate estimate can be applied to a variety of situations.   For the sunspot magnetic field,   the amplification rate  turns out to be in the range of   a few days to years as will be shown below. The sunspot magnetic fields are expected to grow  in the solar convection zone, starting as a small flux tube.  
The pressure balance for such a flux tube can be written as 
\begin{equation}
p_{\mathrm{out}}=p_{\mathrm{in}}+\frac{B^{2}}{ \ 8 \pi},  \label{t-17}
\end{equation}
 and  then, the pressure outside is higher than the  one inside. Since $p=nT$, the density and   temperature in the strong magnetic field region are lower than those in the outside region. This  reasoning seems to agree with the observational fact that solar sunspots look dark  due to cooler temperatures compared to the surrounding areas.   Since the plasma density  inside the sunspot is lower than the outside, there can be  an inward diffusion flow.

 The solar convection zone consists entirely of plasma, and the plasma parameters  vary in a wide range.  When the temperature is chosen as $T = 2.3 \times 10^{6} K \sim 100 eV$ with  the magnetic field as $B=5$ gausses  and  the tube size  about  $L =300$ km,   the magnetic amplification rate becomes roughly as  $\gamma \sim 6.25 \times 10^{-6}  \mathrm{/sec} \sim 0.012  /\mathrm{day} $. The amplification rate depends on the plasma parameters and can vary widely, but at least this example shows how the growth rate estimation can be applied.

Maintaining the flux tube requires an external and internal plasma source and sink, respectively. The ionization and recombination of atoms  can act as source and sink, respectively.  If there is a recycling process of plasma ions,   the density  profile and magnetic fields can be sustained for a much longer time than the particle diffusion time, which is utilized in fusion research machines.    In a controlled thermonuclear fusion device,  recycling of the working gas at the boundary of the plasma is necessary to keep the plasma longer than the particle-confinement times \cite{Marmar78}. It was shown that because of a high recycling rate, the plasma can be sustained for multiple particle-confinement times without the injection of gas from external sources in the vacuum system.  
A similar recycling process can also occur in  other plasmas through ionization and recombination (see Figure \ref{fig1}).

\begin{figure}[h]
\centering
\includegraphics[width=12cm]{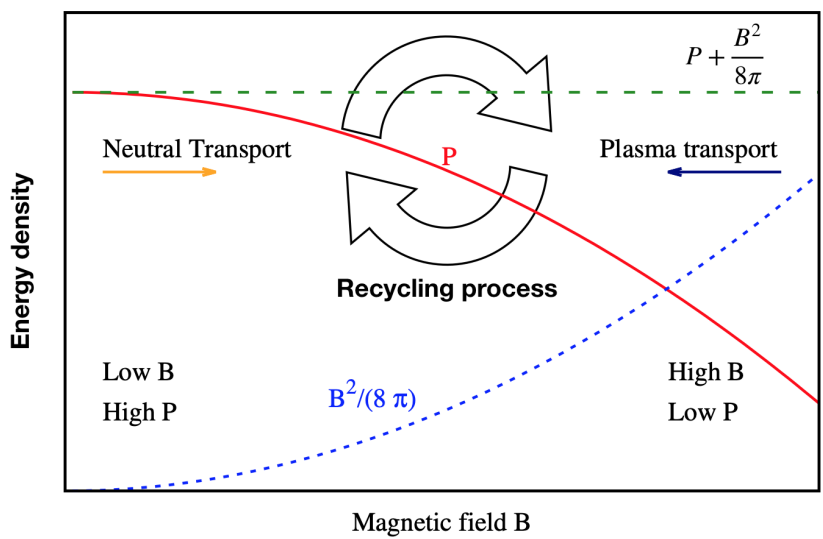}
\caption{Plasma pressure and magnetic field energy density profiles.  The recycling process can occur as shown}
\label{fig1}
\end{figure}

The  transport dynamo mechanism by diffusion can be important in the laboratory plasma.   In  old-day Ohmic discharge  experiments, it was found that a ring shape plasma initially formed near the wall, which quickly filled the center. During this process, magnetic fields of  Bessel-function shapes were shown to be generated   \cite{Lovberg}. This subject will be discussed again later.

\section{Necessity of additional dynamo mechanism in the mean-field dynamo theory }

The mean-field dynamo theory is currently  the main approach in pursuing the dynamo mechanism.   In this theory, the magnetic field is generated by the nonlinear coupling of the flow velocity to the magnetic perturbation.
When the field and velocity variables are decomposed into the mean and fluctuation parts, the induction equation (\ref {7}) can be written as

\begin{equation}
\frac{\partial \overline{\mathbf{B}}}{\partial t}=\nabla \times(\overline{\mathbf{v}} \times \overline{\mathbf{B}})+\nabla \times \mathcal{E}+  \frac{\eta c^{2}}{4 \pi}  \nabla^{2} \overline{\mathbf{B}}, \label{n-1}
\end{equation}
where 
\begin{equation}
\overline{\mathbf{v}^{\prime}}=\overline{\mathbf{B}^{\prime}}=0, \label{n-2}
\end{equation}
and 
\begin{equation}
 \mathcal{E}=\overline{\mathbf{v}^{\prime} \times \mathbf{B}^{\prime}} . \label{n-3}
\end{equation}
 The overline  denotes the average value, and the prime  denotes the fluctuation.
 The electromotive force  can be obtained  from the average of the combined variables of velocity and magnetic fluctuations.
Although the mean-field dynamo theory can explain certain dynamo activity, such as  the solar butterfly diagram, there are still   some concerns left about the mean-field dynamo theory.

The necessity for reconsideration or additional mechanisms to the current dynamo theory  has been proposed by some authors \cite{paul16, Wright16, GK2011}. One of the major drawbacks of  the current dynamo theory pointed out is that it cannot adequately  amplify the seed toroidal field to the desired levels  \cite{Fan2009}.

The active regions of the Sun's surface originate from strong toroidal magnetic fields that lie deep in the solar convection zone.  
Thin flux tube models of emerging flux loops through the solar convective envelope require a strong  toroidal magnetic
field at the base of the solar convection zone. Its magnetic energy is about 10 to 100 times the kinetic energy \cite{Fan2009}.   However, the mean-field dynamo theory has a difficulty in increasing the magnetic energy beyond the equipartition energy.

Ideal MHD is based on various approximations and many effects are omitted from the ideal MHD description. When the electron pressure effect becomes important, the Biermann battery can play an important role.  When the electron pressure term is included,  Ohm's law (\ref{t-3}) becomes
{\begin{equation}
\mathbf{E}+\frac{1}{c} \mathbf{v} \times \mathbf{B}  =  - \frac{1}{nc} \mathbf{\Gamma} \times \mathbf{B}+ \eta \mathbf{J} -  \frac {\nabla {p_e}}{n_e e} , \label{t-3-1}
\end{equation}

and this equation gives rise to the magnetic induction equation as,

{\begin{equation}
\frac{\partial \mathbf{B}}{\partial t}-\nabla \times(\mathbf{v} \times \mathbf{B}) =
 \nabla \times ( {\frac{1}{n} \mathbf{\Gamma} \times \mathbf{B}}) +
  \frac{\eta c^{2}}{4 \pi} \nabla^{2} \mathbf{B}
  -  c\frac{\nabla {n_e} \times  {\nabla {p_e}}}{{n_e}^2 e}. \label{t-4-1}
\end{equation}

The last term in equation (\ref{t-4-1}) is the Biermann battery term.  Recently,  Tzeferacos et al. studied the turbulent dynamo in colliding jet plasma experiments using a high-power laser and could observe the turbulent magnetic generation \cite{Tzeferacos18}. However, Ryu et al. pointed out by using simulation results  that the Biermann battery effects may in fact have  contributed more than they expected to their observed results \cite{Ryu_HED}.   The magnetic energy of a small-scale turbulent dynamo can be significantly affected by the Biermann battery effect. 
Whether the scale of  magnetic generation is large  or small, the search for additional dynamo mechanisms seems to be in order.

\section{Exact solution for magnetic evolution}

If  the resistivity term in equation (\ref{t-4})  is moved to the LHS and the flow velocity is  set to be zero such that $\mathbf{v}=0$, the magnetic induction equation  becomes a simple heat equation with  a source term. 
\begin{equation}
\frac{\partial \mathbf{B} }{\partial t}-\frac{\eta c^{2}}{4 \pi} \nabla^{2} \mathbf{B} =\nabla \times(\frac{1}{n} \mathbf{\Gamma} \times \mathbf{B}). \label{e-2}
\end{equation}

In the absence of a source term, the homogeneous partial differential equations can be solved in cylindrical geometry using the Bessel function.
The complexity of the source term on the RHS  makes  it difficult to solve equations analytically. However,   Lee and Ryu \cite{Ryu_PoP,Ryu_JPD} have shown that
  if the diffusion flow velocity $v_D$ is of 
Bohm-type,  analytically obtaining specific solutions is possible for magnetic fields with axial or azimuthal symmetry.  By using a simple density profile and the Green's function technique,  the analytical solution  for the magnetic field was obtained as follows.


\begin{equation}
\begin{array}{l}
B_{z}(r, t)=B_{z}^{\mathrm{g}}(r, t)+B_{z}^{P}(r, t) \\
\quad=\sum_{m=1}^{\infty}\left[d_{m}+\int_{0}^{t} \mathrm{e}^{a \alpha_{0 x}^{2} t^{\prime} / L^{2}} k_{m}\left(t^{\prime}\right) \mathrm{d} t^{\prime}\right] \mathrm{e}^{-a \alpha_{0 \pi}^{2} t / L^{2}} \\
\quad \times J_{0}\left(\alpha_{0 m} \frac{r}{L}\right), 
\end{array}  \label{e-3}
\end{equation}

\begin{equation}
\begin{array}{l}
B_{\theta}(r, t)=B_{\theta}^{\mathrm{g}}(r, t)+B_{\theta}^{\mathrm{P}}(r, t) \\
\quad=\sum_{m=1}^{\infty}\left[f_{m}+\int_{0}^{t} \mathrm{e}^{a \alpha_{1 m}^{2} t^{\prime} / L^{2}} g_{m}\left(t^{\prime}\right) \mathrm{d} t^{\prime}\right] \mathrm{e}^{-a \alpha_{1}^{2} t / L^{2}} \\
\quad \times J_{1}\left(\alpha_{1 m} \frac{r}{L}\right). 
\end{array} \label{e-4}
\end{equation}

Here, $\alpha_{0 m}$ and $\alpha_{1 m}$ are the $m$th zeroes of Bessel functions, and  $\mathrm{g}$ and $P$ in the superscripts denote homogeneous and particular solutions, respectively.
 These results indicate that Bohm-type diffusion can amplify or reduce the magnetic field in the form of a Bessel function depending on the diffusion direction in the cylindrical approximation.
For  a 100 eV plasma in an apparatus of L=0.1 m with a radially increasing density profile as $ n  \sim exp (-x/L)$, the magnetic amplification in time becomes $(ckT  /{16e L^2 )  \alpha_{0 1}}^2 t$ with  ${\alpha_{0 1}=0.24}$, which becomes  $3.6  \times 10^{7} t $ /sec.
Compared to the growth rate estimated using eq.(\ref{t-16}), the growth rate obtained from the exact solution has an extra factor  ${\alpha_{0 1}}^2$ associated with the first zero of  Bessel function $J_0$ . The amplified magnetic field can be eventually saturated by the reduction of the diffusion coefficient  due to increased magnetic fields and  the decreased temperature by the increased magnetic energy.
  
On the other hand, if the magnetic field has dual directional components ${\mathbf B} = {\mathbf B_z}+ {\mathbf B_{\theta}}$, the  diffusion  source term becomes $ \nabla \times ({\mathbf {v_D }  \times {\mathbf B} }) = \nabla  \times ({\mathbf v_D } \times ({\mathbf B_z}+ {\mathbf B_{\theta}))}$, with the magnitude of the magnetic field  $B = \sqrt{ B_z^2 + B_{\theta}^2}$, which determines the diffusion coefficient. Then, unlike the unidirectional the cases, the diffusion source term is no longer separable for each direction. In this case,  simple growing and damping solutions cannot be obtained,  and  a computational study can only provide solutions for coupled magnetic induction equations. However, such a computational study is  completely lacking at present.


As mentioned previously, many laboratory and astrophysical plasmas exhibit  magnetic fields in the form of  Bessel-functions,  $J_0$ and $J_1$, in the toroidal and poloidal directions, respectively. 
Bessel function shaped magnetic fields have been shown  to occur naturally in the force-free equilibrium, 
 \begin{equation}
 \frac {1}{c} (J \times B)   = 0. \label{e-5}
 \end{equation} 
 The path to this special magnetic configuration  is currently  thought to be due to turbulent relaxation toward the minimum energy state \cite{Taylor1,Taylor2}, but how this relaxation can be achieved by turbulence in a deterministic system is not fully understood.  
   Magnetic field evolution equations shown above in (\ref{e-3}, \ref{e-4}) may provide further insight into this magnetic relaxation.  As described above, if the magnetic evolution is governed by plasma transport,  turbulence can also affect magnetic relaxation via plasma transport.   This particular diffusion effect by turbulence on the magnetic evolution can be investigated in  plasma devices like the Reversed Field Pinch (RFP) or the Field Reversed Configuration (FRC).


\section{Conclusion}


Turbulence plays an important role in the dynamo mechanism.    Turbulence can generate the magnetic field  through the mean-field dynamo mechanism, but  inclusion of cross-field diffusion enables a new kind of dynamo mechanism. 
In this new mechanism, turbulence can generate magnetic fields in the following manner: turbulence induces diffusion of fluid through random motion and then generates a magnetic field through the cross-field diffusion flow. (see Figure \ref{fig2})

\begin{figure}[h]
\centering
\includegraphics[width=12cm]{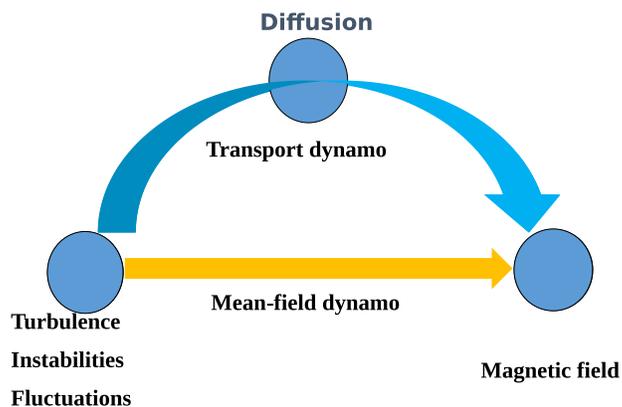}
\caption{Turbulence can generate magnetic fields directly via mean-field dynamo and indirectly via transport dynamo. D denotes the diffusion process}
\label{fig2}
\end{figure}

 Plasma diffusion across the magnetic field has  been observed in many laboratory plasmas.   In thermonuclear fusion plasmas, the cross-field diffusion plays an important role in plasma-confinement and thus has been intensively studied.  However, the potential of  diffusion flow to generate  magnetic fields  has been overlooked.  Despite the fact that significant amounts of magnetic fields can be induced by diffusion, there is no dedicated experiment yet to measure the plasma diffusion in relation to magnetic field generation. 
  
    The generation of a magnetic field by diffusion  has a few distinguishing characteristics: First, since  magnetic generation is  mediated by  a very slow diffusion process, it can  occur  on a much slower time scale  than the Alfvén time scale. Second,  this mechanism can only occur in nonuniform situations,  because there is no diffusion flow in a uniform situation.   In addition, turbulence can greatly enhance the magnetic generation by increasing the diffusion coefficient.  Furthermore, this process is irreversible.  Therefore, the diffusion flow is completely different from the advection flow. The irreversible diffusion flow   has the ability to generate  magnetic fields due to its innate property of having a relative velocity to the magnetic field line. This effect is absent in the advection flow. 
       Therefore, it can be concluded that there are two kinds of flows in  MHD;  advection, where  bulk fluid volume moves  with magnetic field lines at the same velocity, and diffusion, where fluid permeates through the  magnetic field line with a relative velocity. Both satisfy   Ohm's law and flux conservation  required in ideal MHD. 

 Plasma transport by diffusion is a widespread  phenomenon in laboratory and astrophysical plasmas.  Diffusion flows have the ability to generate  a magnetic field. However, at present, confirming evidence is lacking.   If the  magnetic field could be sufficiently  amplified by a transport flow, it could have significant implications for a broad range of magnetic activities in plasma.  From this perspective,  the transport dynamo mechanism  may need more attention and more  theoretical and experimental studies.




\section{Acknowledgements}

This work was supported by IBS (Institute for Basic Science)
under the contract IBS-R012-D1.

\end{document}